\def\paperID{0000}
\def\confName{CVPR}
\def\confYear{2025}

\def\paperTitle{Person-In-Situ: Scene-Consistent Human Image Insertion\\with Occlusion-Aware Pose Control}

\def\authorBlock{
    Shun Masuda%$\thanks{Equal contribution} 
    \qquad
     Yuki Endo %\footnotemark[1]
     \qquad
    Yoshihiro Kanamori \\
    University of Tsukuba
}

\newif\ifreview 
\newif\ifarxiv \newcommand{\arxiv}{\arxivtrue}
\newif\ifcamera 
\newif\ifrebuttal 

\arxiv % \review OR \arxiv OR \cameraready

\pdfoutput=1
\documentclass[10pt,twocolumn,letterpaper]{article}
\ifreview \usepackage[review]{cvpr} \fi
\ifarxiv \usepackage[pagenumbers]{cvpr} \fi
\ifrebuttal \usepackage[rebuttal]{cvpr} \fi
\ifcamera \usepackage{cvpr} \fi

\usepackage{graphicx}	
\usepackage{amsmath}	
\usepackage{amssymb}	
\usepackage{booktabs}
\usepackage{times}
\usepackage{microtype}
\usepackage{epsfig}
\usepackage[table,xcdraw,dvipsnames]{xcolor}
\usepackage{caption}
\usepackage{float}
\usepackage{placeins}
\usepackage{color, colortbl}
\usepackage{stfloats}
\usepackage{enumitem}
\usepackage{tabularx}
\usepackage{xstring}
\usepackage{multirow}
\usepackage{xspace}
\usepackage{url}
\usepackage{subcaption}
\usepackage{xcolor}
\usepackage[hang,flushmargin]{footmisc}

\ifcamera \usepackage[accsupp]{axessibility} \fi

\usepackage{comment}
\usepackage{booktabs}
\usepackage{multirow}
\usepackage{graphicx}
\usepackage{xfrac}
\usepackage{placeins}
\usepackage[T1]{fontenc}
\usepackage{lmodern}
\usepackage{type1cm}
\usepackage{diagbox}
\usepackage{pifont}% http://ctan.org/pkg/pifont

\usepackage{url}

\usepackage{color}
\newcommand{\revA}[1]{#1}
\newcommand{\ykA}[1]{#1}
\newcommand{\smA}[1]{#1}

% ...

\ifarxiv  \fi

\newcommand{\R}[1]{{%
    \textbf{%
        \ifstrequal{#1}{1}{\textcolor{red}{R#1}}{%
        \ifstrequal{#1}{2}{\textcolor{blue}{R#1}}{%
        \ifstrequal{#1}{3}{\textcolor{magenta}{R#1}}{%
        \ifstrequal{#1}{4}{\textcolor{teal}{R#1}}{%
                           \textcolor{cyan}{R#1}%
        }}}}%
    }%
}}

\usepackage{xr-hyper}

\makeatletter
\newcommand*{\addFileDependency}[1]{
  \typeout{(#1)}
  \@addtofilelist{#1}
  \IfFileExists{#1}{}{\typeout{No file #1.}}
}

\makeatother
\newcommand*{\myexternaldocument}[1]{
    \externaldocument{#1}
    \addFileDependency{#1.tex}
    \addFileDependency{#1.aux}
}

\definecolor{cvprblue}{rgb}{0.21,0.49,0.74}
\usepackage[pagebackref,breaklinks,colorlinks,citecolor=cvprblue]{hyperref}
\usepackage[capitalize]{cleveref}
\crefname{section}{Sec.}{Secs.}
\crefname{table}{Table}{Tables}
\crefname{figure}{Fig.}{Figs.}

\ifarxiv \crefname{appendix}{App.}{Apps.}
\else \crefname{appendix}{Suppl.}{Suppls.} \fi

\unless\ifarxiv \myexternaldocument{_supplementary} \fi

\begin{document}
%% TITLE
\title{\paperTitle}
\author{\authorBlock}
%\maketitle
\twocolumn[{%
\renewcommand\twocolumn[1][]{#1}%
\maketitle
\includegraphics[width=1\linewidth]{fig/teaser_new_model.pdf}
\vspace{-2em}
\captionof{figure}{We tackle \ykA{a} novel problem of occlusion-aware human image \ykA{insertion} with \ykA{explicit} pose control, which cannot be handled by \ykA{the state-of-the-art method}~\cite{Affordanceinsertion}. Our method \ykA{can insert} a person in a specified pose at \ykA{an} appropriate depth within a scene, without altering the \ykA{scene's} appearance.\vspace{1em}}
\label{fig:teaser}
}]
\begin{abstract}
% Abstract goes here.
Compositing human \ykA{figure}s into scene images has broad applications in areas such as entertainment and advertising.
However, existing methods often \ykA{cannot handle occlusion of the inserted person by foreground objects and unnaturally place the person} in the frontmost layer. 
Moreover, they offer limited control over the inserted person's pose.
To address these challenges, we propose two methods. Both allow explicit pose control via a 3D body model and leverage latent diffusion models to synthesize the person at a contextually appropriate depth, naturally handling occlusions without requiring occlusion masks.
The first is a two-stage approach: the model first learns a depth map of the scene \ykA{with} the person through supervised \ykA{learning}, \ykA{and} then synthesizes the person accordingly. The second method learns occlusion implicitly and synthesizes the person directly from input data without explicit depth supervision.
Quantitative and qualitative evaluations show that both methods outperform existing approaches by better preserving scene consistency while accurately reflecting occlusions and user-specified poses.
\end{abstract}

\section{Introduction}
\label{sec:Introduction}

The task of human image composition aims to seamlessly integrate a person into a scene while maintaining contextual consistency. 
This \ykA{technique} has diverse applications in areas such as advertising and entertainment.
\ykA{The} state-of-the-art method by Kulal et al.~\cite{Affordanceinsertion} synthesizes a person from another image into a user-specified region of a scene \ykA{with} a natural pose.
\ykA{Although inspiring, their method has several drawbacks.}
\ykA{First}, it does not allow explicit pose control, often leading to unintended results.
\ykA{Secondly, occlusions by} foreground objects also remain difficult to handle.
\ykA{A}ccurate occlusion can be achieved by \ykA{elaborating} detailed masks \ykA{including occluded regions}, \ykA{which} is time-consuming and labor-intensive.
\ykA{Lastly}, the scene appearance within the masked region may be unintentionally altered during synthesis.

\begin{figure}[t]
    \centering
    \includegraphics[width=1.0\linewidth]{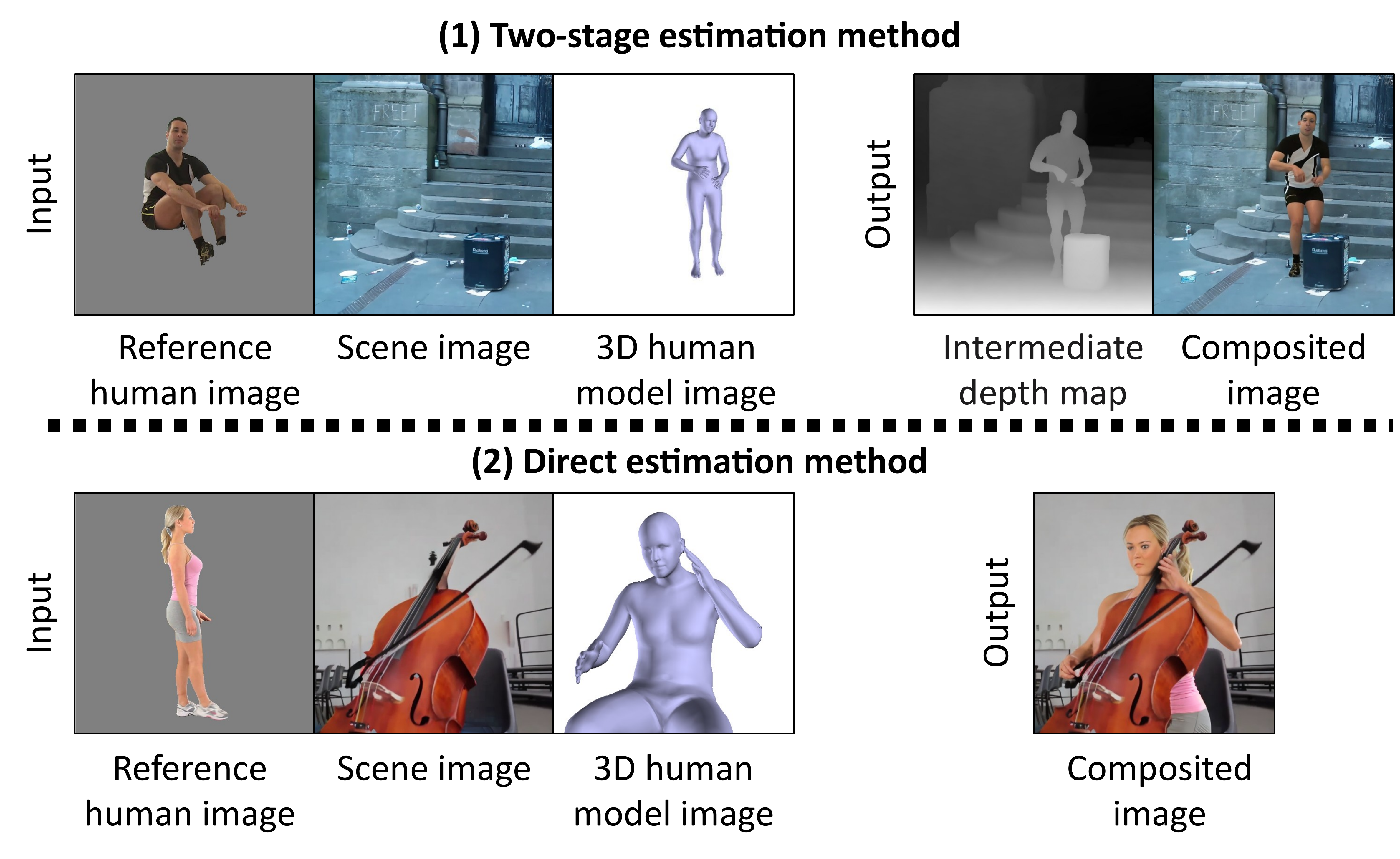}
    \caption{
    \ykA{Our two methods} for human image composition:
    \revA{(1)} a two-stage estimation method, which first estimates an intermediate depth map and then composites the final output; and \revA{(2)} a direct estimation method, which synthesizes the composited image in a single step.
    }
    \label{fig:abst}
\end{figure}

In this paper, we tackle the problem of human image composition that supports
\ykA{occlusion by foreground objects and explicit pose control}
(see Figure~\ref{fig:teaser}).
To this end, we propose two methods. Both methods take as input a reference human image (i.e., the person to be composited), a scene image, and a rendered image of a 3D human model~\cite{SMPL:2015} specifying the target pose.
The 3D model is rendered without occlusion at the desired position and pose within the scene\ykA{, and the 3D model's depth does not have to be consistent with the scene depth.}
Using a latent diffusion model (LDM)~\cite{StableDiffusion}, our approach places the person at an appropriate depth in the scene, enabling occlusion-aware composition without requiring explicit occlusion annotations.
The key difference between the two methods lies in how the scene depth, including the person, is learned either explicitly or implicitly. The first is a two-stage method (Figure~\ref{fig:abst}, top): the first stage explicitly learns a depth map of the scene with the person via supervised \ykA{learning}, and the second stage synthesizes the person based on this map. The second method directly synthesizes the person from the input data (Figure~\ref{fig:abst}, bottom), learning occlusion implicitly without predicting depth. In essence, the two-stage method decomposes the direct method into two subtasks: depth understanding and depth-aware image synthesis.

Our key contributions are summarized as follows:
\begin{enumerate}
    \item \textbf{Occlusion- and pose-aware composition}: We address \ykA{a} novel problem of inserting a person in a specified pose at the correct depth within a scene.
    \item \textbf{Two composition strategies}: We introduce and compare two methods: (1) a two-stage method with intermediate depth prediction, and (2) a direct method that implicitly learns occlusions.
    \item \textbf{Annotation-free training}: Our pipeline automatically generates training data for occlusion learning without manual annotations. 
\end{enumerate}
Quantitative and qualitative evaluations show that our methods outperform \ykA{the state-of-the-art} method by accurately compositing people in specified poses, reproducing occlusions, and preserving surrounding scenes. 
\section{Related Work}
\paragraph{
Human pose editing.}

Several methods have been proposed for editing a person's pose in an image using pose information such as joint positions, 
generating still images~\cite{PoseDDM,PoseLDM,Okuyama_2024_WACV} \ykA{and} videos~\cite{AnimateAnyone,MagicAnimate,Champ}.
However, unlike our work, these methods do not address human composition into a different scene or occlusion by scene objects.

\paragraph{
Object composition.
}

Numerous methods have been proposed for compositing objects specified by prompts or reference images into different scene images. Stable Diffusion~\cite{StableDiffusion}, a latent diffusion model (LDM) trained on large-scale datasets, enables prompt-based inpainting by synthesizing content into masked regions of a scene image. Building on its prior knowledge, several methods allow intuitive image composition using reference images instead of text prompts~\cite{PaintByExample,AnyDoor}. However, these approaches mainly target general object synthesis, not human-centric composition.

A more human-focused method by Kulal et al.~\cite{Affordanceinsertion} adopts a similar learning framework to synthesize people in poses that match scene affordances. In \ykA{their} method, users must manually specify a composition region via a mask. While rough masks are easy to define, they can cause unwanted changes to the scene, whereas detailed masks improve accuracy but are labor-intensive to create.

The method proposed by Lee et al.~\cite{Compose_and_Conquer} handles general object-scene composition using depth maps to control foreground and background placement. However, applying this approach to occlusion-aware human composition requires training data with paired images of the same person before and after occlusion, and such data are difficult and costly to obtain.

In contrast, our method specifies the target pose using a 3D human model, which can be easily generated from an estimated pose. It also enables occlusion-aware synthesis without requiring explicit occlusion annotations. Only occluded (final) person images are needed for training; unoccluded versions are not required.
\begin{figure*}[t]
    \centering
    
    \includegraphics[width=1.\linewidth]{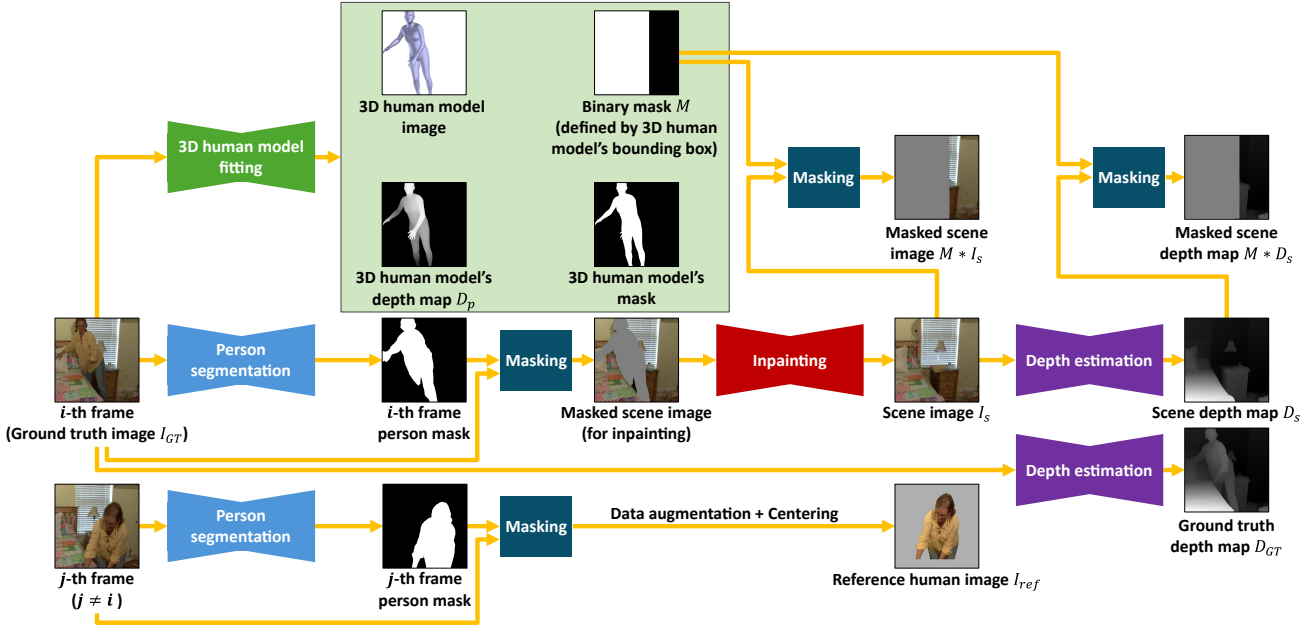}
    \caption{
    Overview of the dataset creation process. Two frames are randomly sampled from a single video: one is used as the reference human image, and the other as the ground-truth image, enabling training with \ykA{paired images (and relevant data)} of the same person in different poses.
    }
    \label{fig:dataset}
\end{figure*}

\section{
Method
}
\label{sec:OurMethod}

This paper aims to achieve occlusion-aware human image composition with explicit pose control. Given a scene image $I_{s}$, 
a \ykA{rendered image of a} 3D human model $I_{p}$,
and a reference human image $I_\mathit{ref}$, our method synthesizes the person with the specified pose at an appropriate depth within the scene. To this end, we train a latent diffusion model (LDM) to learn the spatial relationship between the scene and the person.
The 3D human model image $I_{p}$ is rendered with the \ykA{whole} body visible regardless of occlusion \ykA{in the output image}, 
while \ykA{our networks are} trained \ykA{so that the person is naturally occluded by a foreground object if appropriate}.
To help the LDM capture \ykA{the} front-back relationship \ykA{between the person and scene objects}, we \ykA{input a} depth map $D_{p}$ obtained during rendering instead of $I_{p}$ itself. Additionally, we input \ykA{a} depth map $D_{s}$ of the scene image, estimated by an existing depth estimation model~\cite{depth_anything_v2}. 
\ykA{Furthermore, to specify where to insert the person}, 
we also \ykA{feed a binary mask $M$ defined by the} bounding box of the person region, 
extracted from $I_{p}$.

In this paper, we propose two methods that differ in whether occlusion is handled explicitly or implicitly. The first is a two-stage estimation method that explicitly addresses occlusion by producing a depth map of the scene with the composited person as an intermediate representation. The second is a direct estimation method that handles occlusion implicitly, without intermediate outputs. We begin by describing the dataset construction process used for training, followed by \ykA{a} detailed explanation of each method.

\subsection{
Dataset Preparation
}\label{sec:model1_data}

\ykA{For supervised learning, we require each pair of images (and their relevant data) in which the same person is in different poses (and possibly at different locations) in the same scene.}
\ykA{We construct such a dataset by utilizing large-scale video datasets, following the approach} by Kulal et al.~\cite{Affordanceinsertion}.
The overall dataset construction process is illustrated in Figure~\ref{fig:dataset}. From each video, we extract a pair of frames: one serves as \ykA{a} reference human image, and the other as the ground-truth \ykA{(GT)} image after pose modification.
We \ykA{detect a person by applying} Keypoint R-CNN~\cite{r-cnn} to each frame and crop \ykA{the person's region to} $512 \times 512$ \ykA{pixels}.
Approximately 30 frames are sampled at regular intervals from each video to construct frame pairs.
The main difference from \ykA{the dataset by} Kulal et al. is \ykA{that we include} 3D human models, inpainted scene images, and their corresponding depth maps. \ykA{We explain the detailed procedures as follows.}

\begin{figure*}[t]
    \centering
    \includegraphics[width=0.8\linewidth]{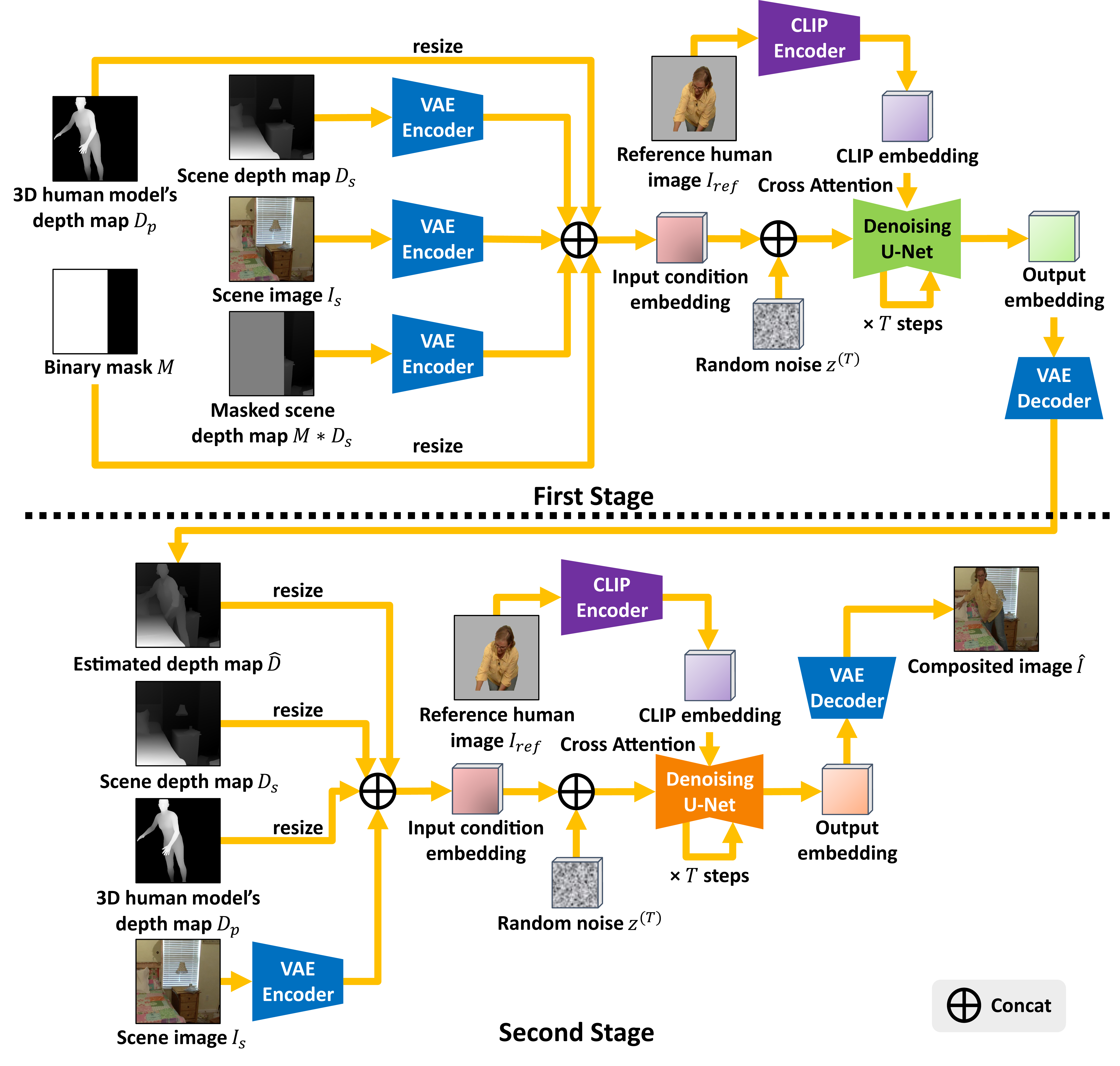}
    
    \caption{
    Network architecture of our two-stage estimation method during inference. In the first stage, the model takes as input the scene image $I_{s}$, reference human image $I_\mathit{ref}$, scene depth map $D_{s}$, 3D human model\ykA{'s} depth map $D_{p}$, 
    \ykA{binary mask $M$ defined by the} 3D human model\ykA{'s} bounding box, 
    and masked scene depth map $\ykA{M \! * \! D_s}$ to predict a depth map $\hat{D}$ of the scene with the person composited. In the second stage, the model uses $I_{s}$, $I_\mathit{ref}$, $D_{s}$, $D_{p}$, and $\hat{D}$ to generate the final composited image.
    }
    \label{fig:proposed_method_B}
\end{figure*}

\paragraph{
Person segmentation.
}

We \ykA{segment out a} person in each frame \ykA{using} Language Segment-Anything~\cite{langSAM}.
First, we detect bounding boxes covering human regions using GroundingDINO~\cite{GroudingDINO}, 
\ykA{and} then \ykA{generate a segmentation mask by applying} Segment-Anything~\cite{SAM} to these regions.
Using this mask, we crop the person\ykA{'s region} to obtain the reference human image $I_\mathit{ref}$. 
\revA{We applied data augmentation to the reference human image $I_\mathit{ref}$ during training, following the baseline method~\cite{Affordanceinsertion}.}
The same mask is also used as the inpainting mask \ykA{for} generating the scene image $I_{s}$.

\paragraph{
3D human model fitting.
}
As the 3D human model, 
we adopt \ykA{a} parametric \ykA{body} model\ykA{, SMPL}~\cite{SMPL:2015}, 
which \ykA{allows us to render} the \ykA{whole} body without occlusion, even when the person is partially occluded in the image. We fit the SMPL model to the person in the ground-truth image $I_\mathit{GT}$ using ProPose~\cite{ProPose}, and use the rendered output as the 3D human model image $I_p$. 
\ykA{We} also obtain its depth map $D_p$ \ykA{during} rendering. 
We normalize the depth values of $D_p$ within [-1, 1], as the 3D model's depth does not have to align with the scene depth in our methods.
Additionally, we generate \ykA{a binary mask $M$ from the person region; we dilate the region of the rendered SMPL model to cover the clothes region and calculate a bounding box of the dilated region to define the mask}.

\paragraph{
Inpainting.
}

We generate a pseudo scene image $I_{s}$ \ykA{(i.e., without any person)} by inpainting the person region in the ground-truth image $I_\mathit{GT}$ \ykA{(i.e., containing the person)} using Stable Diffusion Inpainting v2.0~\cite{StableDiffusion}. Following Lee et al.~\cite{Compose_and_Conquer}, we use the \ykA{same text} prompt ``\textit{empty scenery, highly detailed, no people}'' \ykA{as theirs} for inpainting.

\paragraph{
Depth estimation\ykA{.}
}

We then perform depth estimation on both the scene image $I_{s}$ and the ground-truth image $I_\mathit{GT}$ using Depth Anything~\cite{depth_anything_v2}, obtaining the scene depth map $D_s$ and the ground-truth depth map $D_\mathit{GT}$.

\subsection{
Two-stage Estimation Method}\label{sec:2step_about}

An overview of the two-stage estimation method is shown in Figure~\ref{fig:proposed_method_B}. The first stage explicitly learns a depth map of the scene with the person, while the second stage learns to composite the person image based on this depth.

\begin{figure*}[t]
    \centering
    
    \includegraphics[width=1\linewidth]{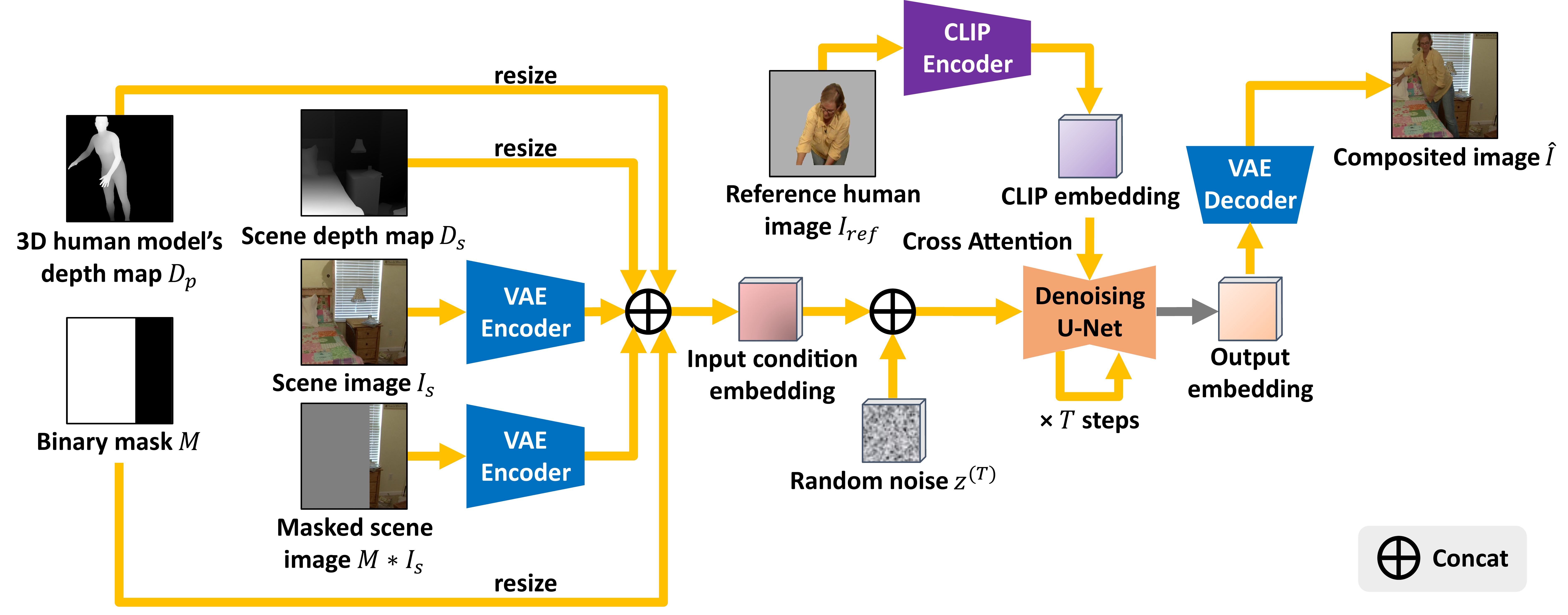}
    \caption{
    Network architecture of the direct estimation method during inference. 
    \ykA{Our method} takes as input the scene image $I_{s}$,
    reference human image $I_\mathit{ref}$, scene depth map $D_{s}$, 3D human model\ykA{'s} depth map $D_{p}$, 
    \ykA{binary mask $M$ defined by the} 3D human model\ykA{'s} bounding box,
    and masked scene image $\ykA{M\!*\!I_s}$.
    \ykA{Our method} then composites $I_\mathit{ref}$, posed according to $D_{p}$, into $I_{s}$ at an appropriate depth.
    }
    \label{fig:proposed_method_A}
\end{figure*}

\paragraph{
Training.}

\ykA{Our first-stage depth estimator leverages powerful generative priors, inspired by a monocular depth estimator, Marigold~\cite{Marigold}. Specifically, we utilize the VAE and U-Net components of the pretrained Stable Diffusion Inpainting v2.0~\cite{StableDiffusion}.}
The ground-truth depth map $D_\mathit{GT}$ is encoded using the VAE encoder $\mathcal{E}$, and Gaussian noise $\epsilon \sim \mathcal{N}(0, \mathbf{I})$ is added to the resulting latent representation according to timestep $t$. The U-Net $\epsilon_\theta$ is then fine-tuned to predict the added noise. The loss function used during training is defined as follows:
\begin{align}
    c^\mathit{2stage}_\mathit{depth} \!\! &= \mathrm{cat}(\mathcal{E}(\ykA{M \! * \! D_s}),\mathcal{E}(I_s),\mathcal{E}(D_s),\mathcal{R}(D_p),\mathcal{R}(\ykA{M})), \\
    c^\mathit{ref} \!\! &=\mathrm{CLIP}(I_\mathit{ref}),  \\
    \mathcal{L}^\mathit{2stage}_\mathit{depth} \!\! &=
        \mathbb{E}
        \Big[ \Vert \epsilon -\epsilon_\theta(\mathrm{cat}(\mathcal{E}(D_\mathit{GT})^{(\mathit{t})},c^\mathit{2stage}_\mathit{depth}),\mathit{t},c^\mathit{ref}) \Vert_{2}^{2}\Big],
    \label{eq:2_1loss}
\end{align}
where \ykA{$*$ denotes element-wise multiplication,} $\mathrm{cat}(\cdot)$ denotes concatenation along the channel dimension, $\mathcal{R}(\cdot)$ resizes an image to match the latent representation dimensions, and $\mathrm{CLIP}(\cdot)$ refers to the CLIP image encoder. The superscript $t$ indicates \ykA{the timestep} in the diffusion process.
In the additional conditioning input $c^\mathit{2stage}_\mathit{depth}$, we apply the VAE encoder to the scene depth map $D_s$, scene image $I_s$, and masked scene depth map $\ykA{M \! * \! D_s}$, following Marigold, to embed them into a shared latent space. Before encoding,
the \ykA{single-channel} depth maps are \ykA{replicated to} three channels.
The number of input channels in the U-Net is adjusted to match the concatenated inputs. The CLIP feature $c^\mathit{ref}$ is \ykA{fed} to the cross-attention layers of the U-Net.
To enable classifier-free guidance (CFG)~\cite{CFG} during inference, we replace the reference human image $I_\mathit{ref}$ with an unconditional image with a probability of 20\% during training. The unconditional image is defined as one filled with the background color of the reference image.

\ykA{The} second stage \ykA{generates} an image from a \ykA{complete} depth map \ykA{(i.e., a depth map of the target person and scene) by also leveraging} the \ykA{generative} prior of Stable Diffusion Inpainting v2.0~\cite{StableDiffusion}. 
Gaussian noise $\epsilon \sim \mathcal{N}(0, \mathbf{I})$ is added to the latent representation of the ground-truth image $I_\mathit{GT}$, obtained via the VAE, according to timestep $t$. The U-Net $\epsilon_\varphi$ is then fine-tuned to predict this noise. The loss function used during training is defined as follows:
\begin{align}
    c^\mathit{2stage}_\mathit{RGB} \!\! &= \mathrm{cat}(\mathcal{E}(I_s),\mathcal{R}(D_\mathit{GT}),\mathcal{R}(D_s),\mathcal{R}(D_p)), \\ 
    \mathcal{L}^\mathit{2stage}_\mathit{RGB} \!\! &= 
        \mathbb{E}\Big[ \Vert \epsilon -\epsilon_	\varphi(\mathrm{cat}(\mathcal{E}(D_\mathit{GT})^{(\mathit{t})},c^\mathit{2stage}_\mathit{RGB}),\mathit{t}, c^\mathit{ref}) \Vert_{2}^{2}\Big].
    \label{eq:2_2loss}
\end{align}
To enable CFG during inference, we replace the reference human image $I_\mathit{ref}$ with an unconditional image with a 20\% probability during training.

\paragraph{
Inference.
}

Figure~\ref{fig:proposed_method_B} illustrates the inference process. In both the first and second stages, random noise $z^{(T)} \sim \mathcal{N}(0, \ykA{\mathbf{I}})$ and conditioning inputs are fed into the U-Net, and denoising is performed over $T$ steps. During this process, CFG ensures that the reference human image $I_\mathit{ref}$ is faithfully reflected in the output. Specifically, in each step of first-stage inference, the predicted noise is updated according to the following equation:
\begin{multline}
\tilde{\epsilon}_\theta(\mathrm{cat}(z^{(\mathit{t})}\!,c^\mathit{2stage}_\mathit{depth}),t,c^\mathit{ref}) = \\
(1 + w^\mathit{2stage}_\mathit{depth}) \, \epsilon_\theta(\mathrm{cat}(z^{(\mathit{t})}\!,c^\mathit{2stage}_\mathit{depth}),t,c^\mathit{ref}) \\
- w^\mathit{2stage}_\mathit{depth} \,
 \epsilon_\theta(\mathrm{cat}(z^{(\mathit{t})}\!,c^\mathit{2stage}_\mathit{depth}),t,\varnothing),
\label{eq:cfg_2_1}
\end{multline}
where $\varnothing$ denotes the CLIP feature of the unconditional image, and $w^\mathit{2stage}_\mathit{depth}$ is the guidance scale. Since the depth map output from the first stage has three channels, we average them to obtain a single-channel depth map, which is then used as the conditioning input for the second stage.
CFG applied during the denoising process in the second stage is defined as follows:
\begin{multline}
\tilde{\epsilon}_\varphi(\mathrm{cat}(z^{(\mathit{t})},c^\mathit{2stage}_\mathit{RGB}),t,c^\mathit{ref}) = \\
 (1+w^\mathit{2stage}_\mathit{RGB}) \, \epsilon_\varphi(\mathrm{cat}(z^{(\mathit{t})},c^\mathit{2stage}_\mathit{RGB}),t,c^\mathit{ref}) \\
 - w^\mathit{2stage}_\mathit{RGB} \, \epsilon_\varphi(\mathrm{cat}(z^{(\mathit{t})},c^\mathit{2stage}_\mathit{RGB}),t,\varnothing),
 \label{eq:cfg_2_2}
\end{multline}
where $w^\mathit{2stage}_\mathit{RGB}$ is a guidance scale.

\subsection{
Direct Estimation Method
}

An overview of the direct estimation method is shown in Figure~\ref{fig:proposed_method_A}. This method composites the reference person into the scene without using any intermediate outputs.

\paragraph{
Training.
}

As in the two-stage estimation method, we fine-tune Stable Diffusion Inpainting v2.0~\cite{StableDiffusion}. The ground-truth image $I_\mathit{GT}$ is passed through the VAE, and Gaussian noise $\epsilon \sim \mathcal{N}(0, \mathbf{I})$ is added to the resulting latent representation \ykA{according to} timestep $t$. The U-Net $\epsilon_\psi$ is then fine-tuned to predict the added noise. The loss function used during training is defined as follows:
\begin{align}
    c^\mathit{direct} \! &= \mathrm{cat}(\mathcal{E}(\ykA{M \! * \! I_s}),\mathcal{E}(I_s),\mathcal{R}(D_s),\mathcal{R}(D_p),\mathcal{R}(\ykA{M})), \\
    \mathcal{L}^\mathit{direct} \! &= \mathbb{E}\Big[ \Vert \epsilon -\epsilon_\psi(\mathrm{cat}(\mathcal{E}(I_\mathit{GT})^{(\mathit{t})},c^\mathit{direct}),\mathit{t},c^\mathit{ref}) \Vert_{2}^{2}\Big].
    \label{eq:1steploss}
\end{align}
To enable CFG during inference, we replace the reference human image $I_\mathit{ref}$ with an unconditional image with a 20\% probability during training.

\paragraph{
Inference.
}

Figure~\ref{fig:proposed_method_A} illustrates the inference process. During inference, random noise $z^{(T)} \sim \mathcal{N}(0, \ykA{\mathbf{I}})$ and conditioning inputs are fed into the U-Net, and denoising is performed over $T$ steps. CFG applied during this process is defined as follows:
\begin{multline}
\tilde{\epsilon}_\psi(\mathrm{cat}(z^{(\mathit{t})},c^\mathit{direct}),t,c^\mathit{ref})  \\ = (1+w^\mathit{direct}) \, \epsilon_\psi(\mathrm{cat}(z^{(\mathit{t})},c^\mathit{direct}),t,c^\mathit{ref}) \\
- w^\mathit{direct} \, \epsilon_\psi(\mathrm{cat}(z^{(\mathit{t})},c^\mathit{direct}),t,\varnothing), \label{eq:cfg_1step}
\end{multline}
where $w^\mathit{direct}$ is the guidance scale.

\section{
Experiments}
\label{sec:Results}

\paragraph{
Experimental settings.
}

\ykA{We implemented our method using Python and the diffusers library~\cite{diffusers}. Each model in our method was trained separately on an NVIDIA RTX A6000 GPU.}
We used 91,424 training samples, 6,329 validation samples, and 1,000 test samples, obtained by preprocessing multiple video datasets~\cite{Dataset_mpii,kinetics_1,hvu,kay2017kineticshumanactionvideo,Charades}. We used the AdamW~\cite{AdamW} optimizer with the initial learning rate of 1e-9, which is linearly increased during the first 10,000 steps and fixed at 5e-5 thereafter. The image resolution was $512 \times 512$, and the batch size was set to 32.
Each model was trained until convergence on the validation set. The direct estimation method was trained for 27 epochs, and the first and second stages of the two-stage estimation method were trained for 26 and 30 epochs, respectively. Training all models took approximately 10 days. During inference, the guidance scale was set to 4.0. The average inference time per image was 11.0 seconds for the two-stage method and 5.6 seconds for the direct method.

\label{sec:quantitative}
\begin{figure*}[t]
    \centering
    
    \includegraphics[width=1.\linewidth]{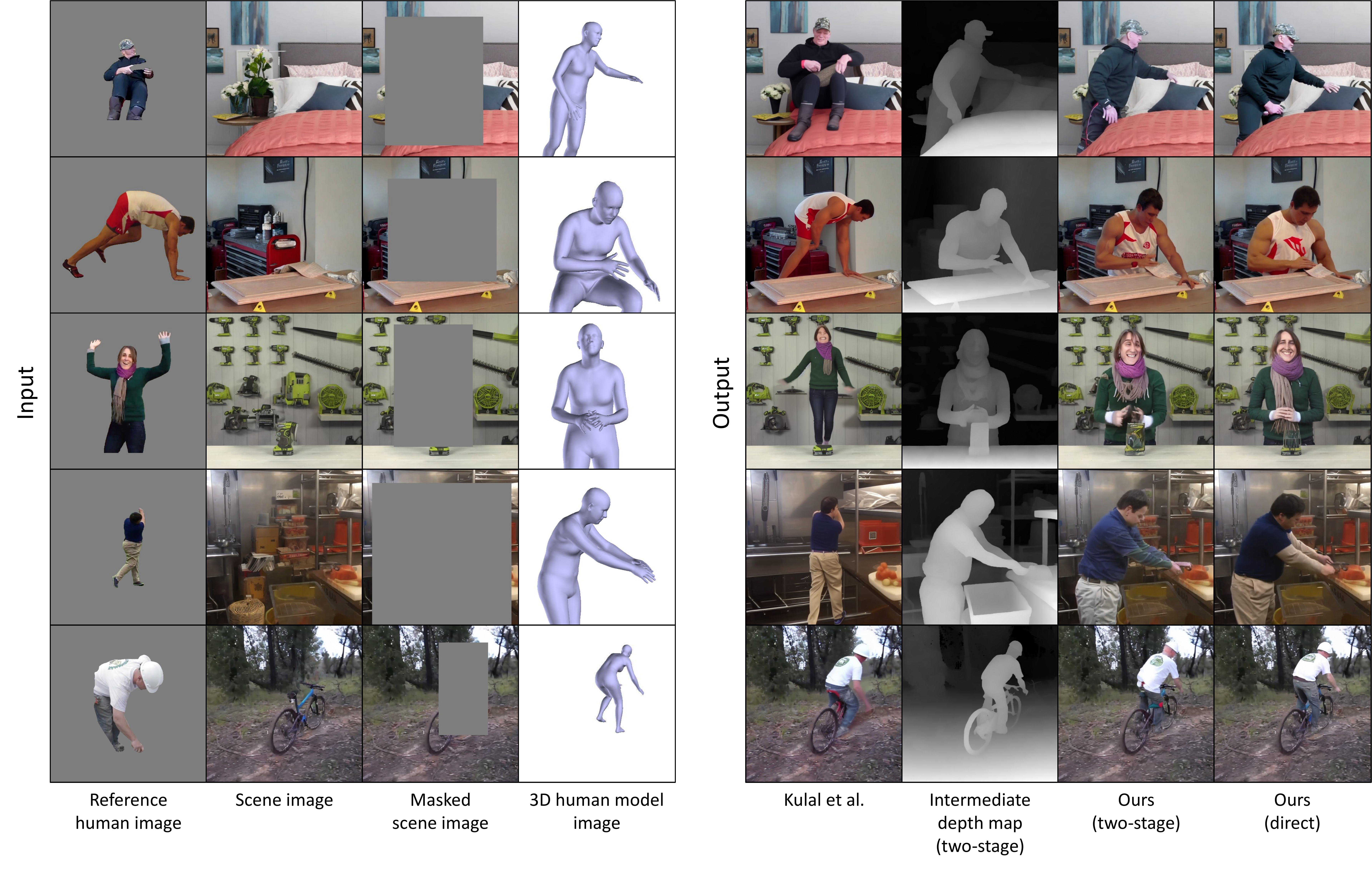}
    \caption{
    Qualitative comparison between our methods and the baseline by Kulal et al.~\cite{Affordanceinsertion}. In addition to a reference human image and a scene image, the baseline uses a mask as input, while our method uses a 3D human model image.
    }
    \label{fig:qualitative}
\end{figure*}

\paragraph{
Compared method.
}

We compare our method with the baseline by Kulal et al.~\cite{Affordanceinsertion}. 
The baseline is fine-tuned using the pre-trained weights of Stable Diffusion (SD) Inpainting v1.5~\cite{StableDiffusion}
(\ykA{see Appendix~\ref{sec:SDVersion} for the quantitative comparisons when our method also uses SD Inpainting v1.5}) by using the same set of video datasets~\cite{Dataset_mpii,kinetics_1,hvu,kay2017kineticshumanactionvideo,Charades} as our method. As described in Section~\ref{sec:model1_data}, approximately 30 frames containing humans are extracted at regular intervals from each video, and the subsequent preprocessing follows \ykA{the} procedure \ykA{of Kulal et al.}
Although the original resolution of \ykA{the} method \ykA{of Kulal et al.} is $256 \times 256$, we train\ykA{ed} and evaluate\ykA{d} it at $512 \times 512$ 
for \ykA{fair comparison} with our method.
The baseline takes as input a masked scene image, a \smA{binary mask}, and a reference human image. 
\ykA{A masked scene image is created by the element-wise multiplication of the scene image and the binary mask.}

Regarding the inference-time inputs, \smA{we use a binary mask created from the bounding box surrounding a person} because of simplicity; a detailed mask helps the baseline specify accurate position and pose, but might require manual labor with explicit consideration of the body shape and occlusion by the foreground object in the scene. Our methods use a 3D human model image as input, whose pose can be easily controlled by manipulating its parameters or obtained via fitting to human images. The user can apply image manipulation to the rendered image of the 3D model without considering the 3D model's occlusion or body shape because occlusion is automatically handled by the network, and the body shape is specified by the reference human image.

\subsection{
Qualitative \ykA{Comparison}
}

Figure~\ref{fig:qualitative} shows the results \ykA{of qualitative comparison}.
\ykA{In} all results, our method clearly \ykA{demonstrates} explicit pose editing using the 3D human model. 
The result in the first row \ykA{shows that our method can} handle occlusions caused by objects such as beds or cushions. Similarly, the results in the second and third rows also highlight this capability.
In the fourth row, we observe that the baseline method introduces noticeable changes to the scene appearance due to rough mask inputs. In contrast, our method preserves the original appearance by using the scene image as input.
In the fifth row, the baseline method fails to preserve the bicycle\ykA{'}s front wheel, as the bounding box of the person includes the front wheel region. In contrast, our method successfully retains the appearance of the front wheel.

We also examine the \ykA{intermediate} depth maps \ykA{predicted in} the two-stage estimation method. All results indicate that the \ykA{intermediate} depth maps faithfully capture the subject's appearance. For instance, in the first row, the reference person's \ykA{cap} is accurately captured \ykA{in} the depth map. Similarly, in the third row, the person's scarf is \ykA{described plausibly}. Furthermore, in all examples, the final composited outputs in the two-stage method align well with the intermediate depth maps.

\begin{figure*}[t]
    \centering
    \includegraphics[width=1.\linewidth]{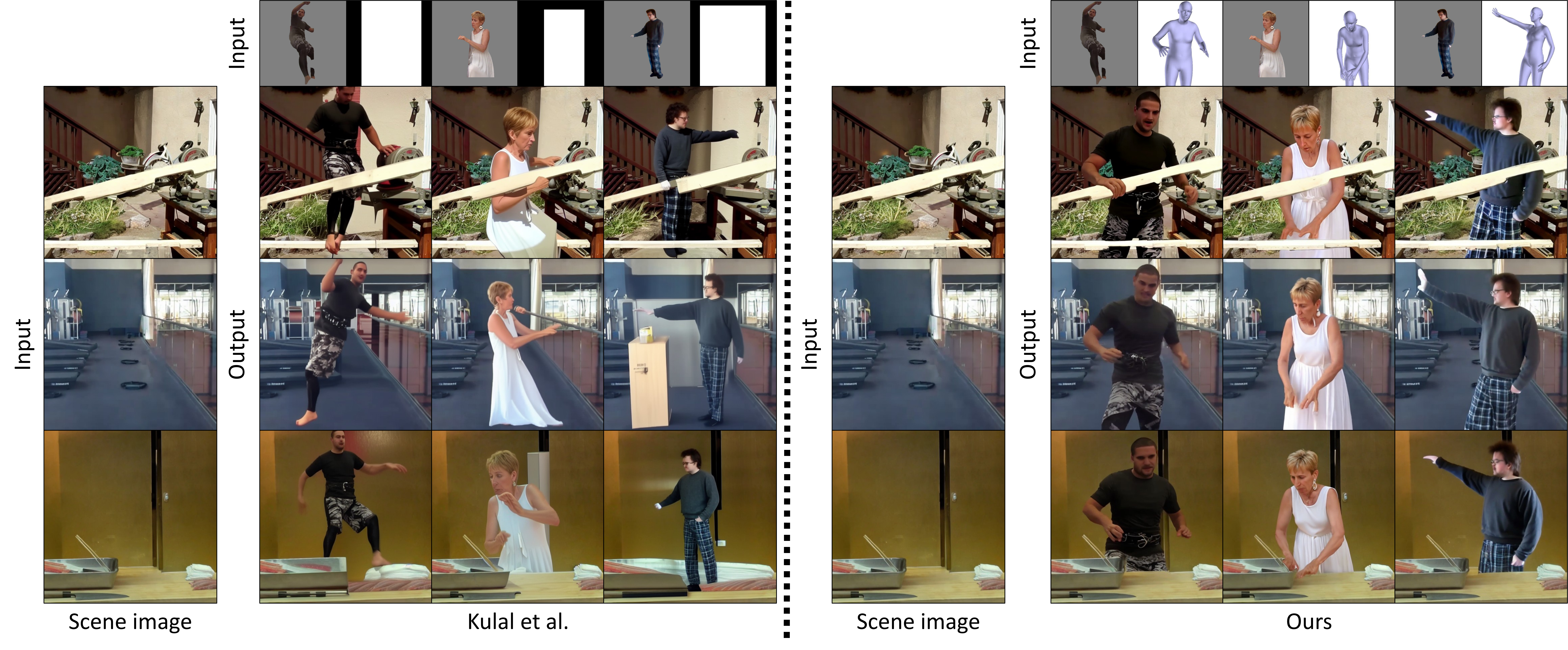}
    \caption{
    Qualitative comparison \ykA{with different scene images and fixed combinations of reference human images and 3D human model images.}
    }
    \label{fig:scene_comparison}
\end{figure*}

\begin{figure*}[t]
    \centering
    \includegraphics[width=1.\linewidth]{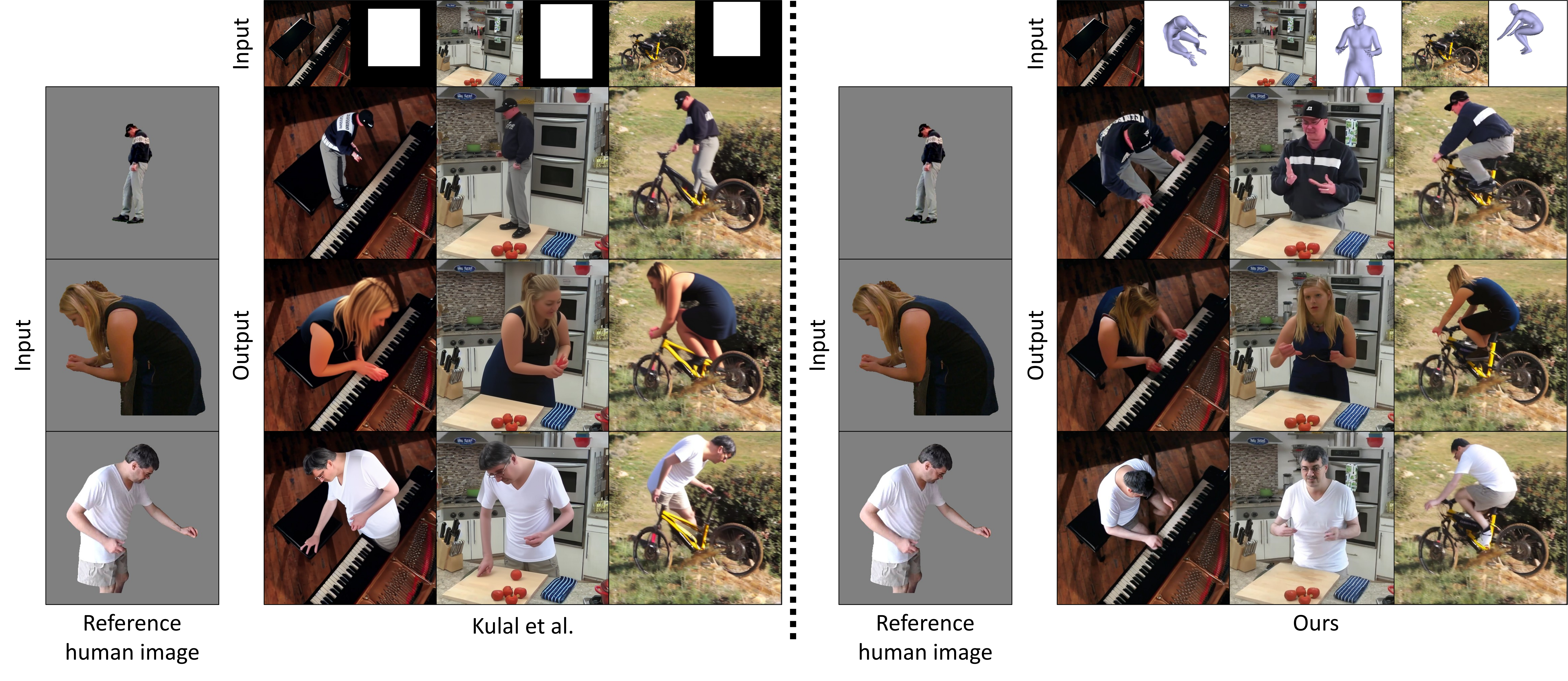}
    \caption{
    Qualitative comparison \ykA{with different reference human images and fixed combinations of scene images and 3D human model images.}    
    }
    \label{fig:person_comparison}
\end{figure*}

\begin{figure*}[t]
    \centering
    \includegraphics[width=1.\linewidth]{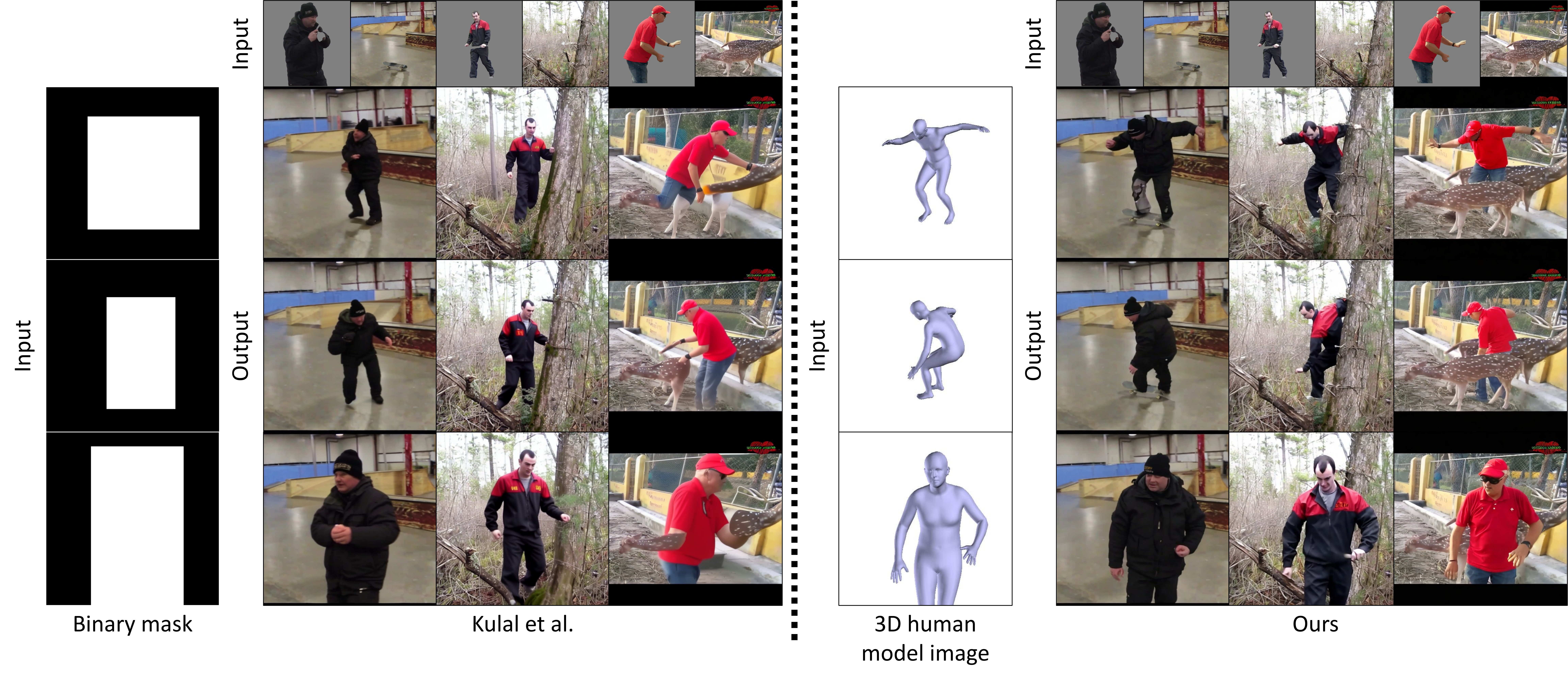}
    \caption{
    Qualitative comparison \ykA{with different 3D human model images and fixed combinations of reference human images and scene images.}    
    }
    \label{fig:pose_comparison}
\end{figure*}

\subsection{
\ykA{Qualitative Comparison with Different Input Combinations}
}

We conducted a qualitative comparison with different combinations of input data: i.e., the scene image, reference human image, and 3D human model image. In the following, we observe generated images by fixing the combination of two of these three inputs and varying the one remaining input. 
Here we used the direct estimation method, which is more accurate than the two-stage variant.

\ykA{Figure~\ref{fig:scene_comparison} shows the result with different scene images and fixed combinations of reference human images and 3D human model images. The masked scene images vary for different binary masks. In the baseline's results, the composited people are in strange poses at inappropriate depths. In contrast, our results show that the people are consistently placed at appropriate depths with appropriate occlusion in specified poses. Quite simply, our method can appropriately accommodate different scenes while the baseline cannot.}

\ykA{Figure~\ref{fig:person_comparison} shows the result with different reference human images and fixed combinations of scene images and 3D human model images. Each column of the baseline's results reveals that the sizes and poses of the composited people largely depend on those of the reference human images. In our results, the sizes and poses are independent of those of the reference human images and well controlled by the 3D human model images. We can also observe that the body shapes of the reference human images are retained after composition, although we use the same 3D body model.}

\ykA{Figure~\ref{fig:pose_comparison} shows the result with different 3D human model images and fixed combinations of reference human images and scene images. In the second column of the baseline's results, the difference in the composited person's depths is not large, which indicates that specifying depth only with masks is difficult. By contrast, our results naturally reflect the person's depths according to the 3D human model's sizes. The third column of the baseline's results exhibits that larger masks cause larger unwanted scene changes, while our method is unlikely to alter the original scene even when the person occupies a large portion of the scene.}

\subsection{
Quantitative Evaluation
}

We conducted a quantitative \ykA{comparison using MSE, SSIM, and CLIP similarity~\cite{CLIP} as the evaluation metrics.}
\ykA{To compare the resultant scene structures, we} also \ykA{evaluated depth maps predicted from the composited images using DepthAnything~\cite{depth_anything_v2}, and used} SSIM and MSE \ykA{as the evaluation metrics}.

As shown in Table~\ref{tab:quantitative_rgb}, 
both \ykA{our direct and two-stage methods}
outperform the baseline across all metrics. Compared to the baseline, our methods generate images that match the ground truth \ykA{more closely}. Among the two, the direct estimation method achieves the best overall performance.

\begin{table}[t]
\setlength{\belowcaptionskip}{10pt}
    \centering
    \caption{
    Quantitative comparison \ykA{of the composited results}.
    The best score for each metric is shown in \textbf{bold}, and the second-best is \underline{underlined}.
    }
    \small
    \begin{tabular}{lccc} \hline
    Method & SSIM$\uparrow$ & MSE$\downarrow$ & CLIP similarity$\uparrow$ \\ \hline \hline
    Kulal et al.~\cite{Affordanceinsertion}& 0.681 & 0.0319 & 0.854 \\  \hline 
         Ours (two-stage) &  \underline{0.710} & \textbf{0.0176} & \underline{0.881}\\
         Ours (direct)&\textbf{0.723} & \underline{0.0177} & \textbf{0.893} \\ \hline 
    \end{tabular}
    \label{tab:quantitative_rgb}
\end{table}

\begin{table}[t]
    \centering
    \setlength{\belowcaptionskip}{10pt}
    \caption{
    Quantitative comparison \ykA{of depth maps predicted from the composited results}.
    }
    \begin{tabular}{lcc} \hline
        Method & SSIM$\uparrow$ & MSE$\downarrow$ \\ \hline \hline
        Kulal et al.~\cite{Affordanceinsertion}& 0.833 & 0.0315 \\  \hline 
         Ours (two-stage) &  \underline{0.880} &  \underline{0.0200} \\
         Ours (direct) &\textbf{0.896} & \textbf{0.0141}  \\ \hline 
    \end{tabular}
    \label{tab:quantitative_depth}
\end{table}

\ykA{Table~\ref{tab:quantitative_depth} shows the quantitative evaluation of depth maps predicted from the composited results.}
The results indicate that both \ykA{our} direct and two-stage methods outperform the baseline across all metrics. Compared to the baseline, our methods place the person at depths more closely aligned with the ground-truth depth maps. Among our methods, the direct estimation approach achieves the best performance.

\section{
Conclusion}

In this paper, we \ykA{have} proposed \ykA{scene-consistent} human image \ykA{insertion} methods that enable explicit pose \ykA{control} and \ykA{account} for occlusions caused by foreground objects in the scene. 
By leveraging \ykA{a 3D human model with full-body information and a pseudo-scene image obtained via inpainting},
\ykA{our networks were} trained without requiring explicit occlusion annotations.
As a result, 
occlusion \ykA{can} be handled through a latent diffusion model.
\ykA{We proposed two variants: i) a two-stage estimation method, in which the first stage estimates an intermediate depth map and the second stage generates an output image based on the intermediate depth map, and ii) a direct estimation method, which directly generates an output image without depth prediction.}
\ykA{Our e}xperimental results demonstrated that both \ykA{our} methods can synthesize \ykA{realistic} images that reflect occlusion by foreground objects.

\ykA{Our methods have several limitations and room for improvement. First, some low-quality training data degrade the accuracy of our methods. We plan to improve our dataset using more sophisticated methods for preprocessing. Second, the detailed appearances in the reference human images, in particular, faces,} are sometimes not sufficiently \ykA{reproduced. This issue is common in the baseline~\cite{Affordanceinsertion}, as both the baseline and ours} use \ykA{the} CLIP image encoder.
Replacing the CLIP encoder with a model like DINOv2~\cite{DINOv2}, which \ykA{can capture} more detailed visual features, \ykA{might alleviate} this problem.

{\small
\bibliographystyle{ieeenat_fullname}
\bibliography{11_references}

\begin{thebibliography}{29}
\providecommand{\natexlab}[1]{#1}
\providecommand{\url}[1]{\texttt{#1}}
\expandafter\ifx\csname urlstyle\endcsname\relax
  \providecommand{\doi}[1]{doi: #1}\else
  \providecommand{\doi}{doi: \begingroup \urlstyle{rm}\Url}\fi

\bibitem[Andriluka et~al.(2014)Andriluka, Pishchulin, Gehler, and
  Schiele]{Dataset_mpii}
Mykhaylo Andriluka, Leonid Pishchulin, Peter~V. Gehler, and Bernt Schiele.
\newblock {2D Human Pose Estimation: New Benchmark and State of the Art
  Analysis}.
\newblock In \emph{{CVPR}}, pages 3686--3693, 2014.

\bibitem[Bhunia et~al.(2023)Bhunia, Khan, Cholakkal, Anwer, Laaksonen, Shah,
  and Khan]{PoseDDM}
Ankan~Kumar Bhunia, Salman~H. Khan, Hisham Cholakkal, Rao~Muhammad Anwer, Jorma
  Laaksonen, Mubarak Shah, and Fahad~Shahbaz Khan.
\newblock {Person Image Synthesis via Denoising Diffusion Model}.
\newblock In \emph{{CVPR}}, pages 5968--5976, 2023.

\bibitem[Carreira et~al.(2022)Carreira, Noland, Hillier, and
  Zisserman]{kinetics_1}
Joao Carreira, Eric Noland, Chloe Hillier, and Andrew Zisserman.
\newblock {A Short Note on the Kinetics-700 Human Action Dataset}.
\newblock \emph{arXiv:1907.06987}, 2022.

\bibitem[Chen et~al.(2024)Chen, Huang, Liu, Shen, Zhao, and Zhao]{AnyDoor}
Xi Chen, Lianghua Huang, Yu Liu, Yujun Shen, Deli Zhao, and Hengshuang Zhao.
\newblock {AnyDoor: Zero-shot Object-level Image Customization}.
\newblock In \emph{{CVPR}}, pages 6593--6602, 2024.

\bibitem[Diba et~al.(2020)Diba, Fayyaz, Sharma, Paluri, Gall, Stiefelhagen, and
  Van~Gool]{hvu}
Ali Diba, Mohsen Fayyaz, Vivek Sharma, Manohar Paluri, J{\"u}rgen Gall, Rainer
  Stiefelhagen, and Luc Van~Gool.
\newblock {Large Scale Holistic Video Understanding}.
\newblock In \emph{European Conference on Computer Vision}, pages 593--610,
  2020.

\bibitem[Fang et~al.(2023)Fang, Chen, Fan, Shuai, Li, and Zhang]{ProPose}
Qi Fang, Kang Chen, Yinghui Fan, Qing Shuai, Jiefeng Li, and Weidong Zhang.
\newblock {Learning Analytical Posterior Probability for Human Mesh Recovery}.
\newblock In \emph{{CVPR}}, pages 8781--8791, 2023.

\bibitem[Han et~al.(2023)Han, Zhu, Deng, Song, and Xiang]{PoseLDM}
Xiao Han, Xiatian Zhu, Jiankang Deng, Yi{-}Zhe Song, and Tao Xiang.
\newblock {Controllable Person Image Synthesis with Pose-Constrained Latent
  Diffusion}.
\newblock In \emph{{ICCV}}, pages 22711--22720, 2023.

\bibitem[He et~al.(2017)He, Gkioxari, Doll{\'{a}}r, and Girshick]{r-cnn}
Kaiming He, Georgia Gkioxari, Piotr Doll{\'{a}}r, and Ross~B. Girshick.
\newblock {Mask {R-CNN}}.
\newblock In \emph{{ICCV}}, pages 2980--2988, 2017.

\bibitem[Ho and Salimans(2022)]{CFG}
Jonathan Ho and Tim Salimans.
\newblock {Classifier-Free Diffusion Guidance}.
\newblock \emph{arXiv:2207.12598}, 2022.

\bibitem[Hu(2024)]{AnimateAnyone}
Li Hu.
\newblock {Animate Anyone: Consistent and Controllable Image-to-Video Synthesis
  for Character Animation}.
\newblock In \emph{{CVPR}}, pages 8153--8163, 2024.

\bibitem[Kay et~al.(2017)Kay, Carreira, Simonyan, Zhang, Hillier,
  Vijayanarasimhan, Viola, Green, Back, Natsev, Suleyman, and
  Zisserman]{kay2017kineticshumanactionvideo}
Will Kay, Joao Carreira, Karen Simonyan, Brian Zhang, Chloe Hillier, Sudheendra
  Vijayanarasimhan, Fabio Viola, Tim Green, Trevor Back, Paul Natsev, Mustafa
  Suleyman, and Andrew Zisserman.
\newblock {The Kinetics Human Action Video Dataset}.
\newblock \emph{arXiv:1705.06950}, 2017.

\bibitem[Ke et~al.(2024)Ke, Obukhov, Huang, Metzger, Daudt, and
  Schindler]{Marigold}
Bingxin Ke, Anton Obukhov, Shengyu Huang, Nando Metzger, Rodrigo~Caye Daudt,
  and Konrad Schindler.
\newblock {Repurposing Diffusion-Based Image Generators for Monocular Depth
  Estimation}.
\newblock In \emph{{CVPR}}, pages 9492--9502, 2024.

\bibitem[Kirillov et~al.(2023)Kirillov, Mintun, Ravi, Mao, Rolland, Gustafson,
  Xiao, Whitehead, Berg, Lo, Doll{\'a}r, and Girshick]{SAM}
Alexander Kirillov, Eric Mintun, Nikhila Ravi, Hanzi Mao, Chloe Rolland, Laura
  Gustafson, Tete Xiao, Spencer Whitehead, Alexander~C. Berg, Wan-Yen Lo, Piotr
  Doll{\'a}r, and Ross Girshick.
\newblock {Segment Anything}.
\newblock \emph{arXiv:2304.02643}, 2023.

\bibitem[Kulal et~al.(2023)Kulal, Brooks, Aiken, Wu, Yang, Lu, Efros, and
  Singh]{Affordanceinsertion}
Sumith Kulal, Tim Brooks, Alex Aiken, Jiajun Wu, Jimei Yang, Jingwan Lu,
  Alexei~A. Efros, and Krishna~Kumar Singh.
\newblock {Putting People in Their Place: Affordance-Aware Human Insertion into
  Scenes}.
\newblock In \emph{{CVPR}}, pages 17089--17099, 2023.

\bibitem[Lee et~al.(2024)Lee, Cho, Yoo, Kim, and Jeong]{Compose_and_Conquer}
Jonghyun Lee, Hansam Cho, Young~Joon Yoo, Seoung~Bum Kim, and Yonghyun Jeong.
\newblock {Compose and Conquer: Diffusion-Based 3D Depth Aware Composable Image
  Synthesis}.
\newblock In \emph{{ICLR}}, 2024.

\bibitem[Liu et~al.(2024)Liu, Zeng, Ren, Li, Zhang, Yang, Jiang, Li, Yang, Su,
  Zhu, and Zhang]{GroudingDINO}
Shilong Liu, Zhaoyang Zeng, Tianhe Ren, Feng Li, Hao Zhang, Jie Yang, Qing
  Jiang, Chunyuan Li, Jianwei Yang, Hang Su, Jun Zhu, and Lei Zhang.
\newblock {Grounding {DINO:} Marrying {DINO} with Grounded Pre-training for
  Open-Set Object Detection}.
\newblock In \emph{{ECCV}}, pages 38--55, 2024.

\bibitem[Loper et~al.(2015)Loper, Mahmood, Romero, Pons-Moll, and
  Black]{SMPL:2015}
Matthew Loper, Naureen Mahmood, Javier Romero, Gerard Pons-Moll, and Michael~J.
  Black.
\newblock {{SMPL}: A Skinned Multi-Person Linear Model}.
\newblock \emph{ACM Trans. Graphics (Proc. SIGGRAPH Asia)}, \penalty0
  (6):\penalty0 248:1--248:16, 2015.

\bibitem[Loshchilov and Hutter(2019)]{AdamW}
Ilya Loshchilov and Frank Hutter.
\newblock {Decoupled Weight Decay Regularization}.
\newblock In \emph{{ICLR} 2019}. OpenReview.net, 2019.

\bibitem[luca medeiros()]{langSAM}
luca medeiros.
\newblock {Language Segment-Anything}.
\newblock \url{https://github.com/luca-medeiros/lang-segment-anything}
  (Accessed on 8/19/2024).

\bibitem[Okuyama et~al.(2024)Okuyama, Endo, and Kanamori]{Okuyama_2024_WACV}
Yuta Okuyama, Yuki Endo, and Yoshihiro Kanamori.
\newblock {{DiffBody}: {D}iffusion-{B}ased {P}ose and {S}hape {E}diting of
  {H}uman {I}mages}.
\newblock In \emph{WACV}, pages 6333--6342, 2024.

\bibitem[Oquab et~al.()Oquab, Darcet, Moutakanni, Vo, Szafraniec, Khalidov,
  Fernandez, Haziza, Massa, El{-}Nouby, Assran, Ballas, Galuba, Howes, Huang,
  Li, Misra, Rabbat, Sharma, Synnaeve, Xu, J{\'{e}}gou, Mairal, Labatut,
  Joulin, and Bojanowski]{DINOv2}
Maxime Oquab, Timoth{\'{e}}e Darcet, Th{\'{e}}o Moutakanni, Huy~V. Vo, Marc
  Szafraniec, Vasil Khalidov, Pierre Fernandez, Daniel Haziza, Francisco Massa,
  Alaaeldin El{-}Nouby, Mido Assran, Nicolas Ballas, Wojciech Galuba, Russell
  Howes, Po{-}Yao Huang, Shang{-}Wen Li, Ishan Misra, Michael Rabbat, Vasu
  Sharma, Gabriel Synnaeve, Hu Xu, Herv{\'{e}} J{\'{e}}gou, Julien Mairal,
  Patrick Labatut, Armand Joulin, and Piotr Bojanowski.
\newblock {DINOv2: Learning Robust Visual Features without Supervision}.
\newblock \emph{Trans. Mach. Learn. Res.}, 2024.

\bibitem[Radford et~al.(2021)Radford, Kim, Hallacy, Ramesh, Goh, Agarwal,
  Sastry, Askell, Mishkin, Clark, Krueger, and Sutskever]{CLIP}
Alec Radford, Jong~Wook Kim, Chris Hallacy, Aditya Ramesh, Gabriel Goh,
  Sandhini Agarwal, Girish Sastry, Amanda Askell, Pamela Mishkin, Jack Clark,
  Gretchen Krueger, and Ilya Sutskever.
\newblock {Learning Transferable Visual Models From Natural Language
  Supervision}.
\newblock In \emph{{ICML}}, pages 8748--8763, 2021.

\bibitem[Rombach et~al.(2022)Rombach, Blattmann, Lorenz, Esser, and
  Ommer]{StableDiffusion}
Robin Rombach, Andreas Blattmann, Dominik Lorenz, Patrick Esser, and
  Bj{\"{o}}rn Ommer.
\newblock {High-Resolution Image Synthesis with Latent Diffusion Models}.
\newblock In \emph{{CVPR}}, pages 10674--10685, 2022.

\bibitem[Sigurdsson et~al.(2016)Sigurdsson, Varol, Wang, Farhadi, Laptev, and
  Gupta]{Charades}
Gunnar~A. Sigurdsson, G{\"{u}}l Varol, Xiaolong Wang, Ali Farhadi, Ivan Laptev,
  and Abhinav Gupta.
\newblock {Hollywood in Homes: Crowdsourcing Data Collection for Activity
  Understanding}.
\newblock In \emph{{ECCV} {(1)}}, pages 510--526, 2016.

\bibitem[von Platen et~al.(2022)von Platen, Patil, Lozhkov, Cuenca, Lambert,
  Rasul, Davaadorj, Nair, Paul, Berman, Xu, Liu, and Wolf]{diffusers}
Patrick von Platen, Suraj Patil, Anton Lozhkov, Pedro Cuenca, Nathan Lambert,
  Kashif Rasul, Mishig Davaadorj, Dhruv Nair, Sayak Paul, William Berman, Yiyi
  Xu, Steven Liu, and Thomas Wolf.
\newblock {Diffusers: State-of-the-art diffusion models}.
\newblock \url{https://github.com/huggingface/diffusers}(Accessed on
  10/4/2024), 2022.

\bibitem[Xu et~al.(2024)Xu, Zhang, Liew, Yan, Liu, Zhang, Feng, and
  Shou]{MagicAnimate}
Zhongcong Xu, Jianfeng Zhang, Jun~Hao Liew, Hanshu Yan, Jia{-}Wei Liu, Chenxu
  Zhang, Jiashi Feng, and Mike~Zheng Shou.
\newblock {MagicAnimate: Temporally Consistent Human Image Animation using
  Diffusion Model}.
\newblock In \emph{{CVPR}}, pages 1481--1490, 2024.

\bibitem[Yang et~al.(2023)Yang, Gu, Zhang, Zhang, Chen, Sun, Chen, and
  Wen]{PaintByExample}
Binxin Yang, Shuyang Gu, Bo Zhang, Ting Zhang, Xuejin Chen, Xiaoyan Sun, Dong
  Chen, and Fang Wen.
\newblock {Paint by Example: Exemplar-based Image Editing with Diffusion
  Models}.
\newblock In \emph{{CVPR}}, pages 18381--18391, 2023.

\bibitem[Yang et~al.(2024)Yang, Kang, Huang, Zhao, Xu, Feng, and
  Zhao]{depth_anything_v2}
Lihe Yang, Bingyi Kang, Zilong Huang, Zhen Zhao, Xiaogang Xu, Jiashi Feng, and
  Hengshuang Zhao.
\newblock {Depth Anything {V2}}.
\newblock In \emph{NeurIPS}, 2024.

\bibitem[Zhu et~al.(2024)Zhu, Chen, Dai, Dong, Xu, Cao, Yao, Zhu, and
  Zhu]{Champ}
Shenhao Zhu, Junming~Leo Chen, Zuozhuo Dai, Zilong Dong, Yinghui Xu, Xun Cao,
  Yao Yao, Hao Zhu, and Siyu Zhu.
\newblock {Champ: Controllable and Consistent Human Image Animation with 3D
  Parametric Guidance}.
\newblock In \emph{{ECCV}}, pages 145--162, 2024.

\end{thebibliography}
}

\ifarxiv \clearpage \appendix \section*{Appendix}
\section{
\revA{Impact of Pre-trained Model Choice}
}
\label{sec:SDVersion}

\revA{To ensure a fair comparison with the baseline~\cite{Affordanceinsertion} that uses Stable Diffusion Inpainting v1.5~\cite{StableDiffusion}, we additionally evaluate our method using the same pre-trained weights. }

For this comparison, we used \ykA{our} direct estimation \ykA{method}, which is more quantitatively and qualitatively effective \ykA{than our two-stage} method.

As shown in Table\ykA{s}~\ref{tab:ablation_rgb} and \ykA{\ref{tab:ablation_depth}},
it is evident that even when using the same weights as the baseline method, \ykA{our} method outperforms the baseline. This confirms that the training approach of \ykA{our} method is effective for the current task.

\begin{table}[h]
\setlength{\belowcaptionskip}{10pt}
    \centering
    \caption{
    Quantitative comparison \ykA{of} the generated results \ykA{using the same SD model weights}.
    }
    \small
    \begin{tabular}{lccc} \hline
    Method & SSIM$\uparrow$ & MSE$\downarrow$ & CLIP similarity$\uparrow$ \\ \hline \hline
    Kulal et al.~\cite{Affordanceinsertion}& 0.681 & 0.0319 & 0.854 \\  \hline 
         Ours (SD v1.5) &  \textbf{0.723} & \textbf{0.0174} & \textbf{0.883}\\ \hline
    \end{tabular}
    \label{tab:ablation_rgb}
\end{table}

\begin{table}[h]
    \centering
    \setlength{\belowcaptionskip}{10pt}
    \caption{
    Quantitative comparison \ykA{of} the depth maps \ykA{predicted from} the generated results \ykA{using the same SD model weights}.
    }
    \begin{tabular}{lcc} \hline
        Method & SSIM$\uparrow$ & MSE$\downarrow$ \\ \hline \hline
        Kulal et al.~\cite{Affordanceinsertion}& 0.833 & 0.0315 \\  \hline 
         Ours (SD v1.5) &\textbf{0.893} & \textbf{0.0144}  \\ \hline 
    \end{tabular}
    \label{tab:ablation_depth}
\end{table} \fi

\end{document}